# The slow relaxation of non-equilibrium state in metal target excited by picosecond electron beam: interferometric and simulation studies

**Barakhvostov S V[1], Volkov N B[1,*], Lipchak A I[1], Tarakanov V P[2],**

**Tkachenko S I[3], Turmyshev I S[1] and Yalovets AP[4]**

[1]Institute of Electrophysics, Russian Academy of Science, Ural Branch,
106 Amundsen Street, Yekaterinburg 620016, Russia
[2]Join Institute for High Temperatures, RussianAcademy of Science,
13/19 Izhorskaya Street, Moscow 127412, Russia
[3]Moscow Institute of Physics and Technology (State University),
9 Instituskiy per., Dolgoprudny, Moscow Region 141701, Russia
[4]South Ural StateUniversity, 76 Lenin Avenue, Chelyabinsk 454080, Russia

*nbv@iep.uran.ru

**Abstract.** The experimental EXCITOR setup for obtaining the intense electron beams of picosecond duration is presented in this work. Cathodes of tungsten, graphite, copper and samarium of the same shape were used in these experiments. The interelectrode gap varied from 0.5 to 6mm. Interferometric technique to detect anode rear side displacement was used. The time evolution of the displacement and speed of the rear side of a copper anode of 0.1 and 3.0 mm thickness was experimentally obtained. Simulation of the space-time characteristics evolution of the anode material is presented.

## 1. Introduction

Powerful beams of charged particles and laser radiation are widely used for studying the physical properties of condensed matter at high energy densities (e. g., see the proceedings of international conferences [1-3] and articles cited in them). First of all the energy absorption of a powerful laser radiation or the current of the fast electrons causes heating of the electronic component of a condensed matter. While the heating of the ion component is a result of the electron-phonon interaction [4]. The time of electron-ion relaxation with respect to energy in a condensed matter is $\tau_\varepsilon \sim 10^{-12}$ s, while the time of establishment of local thermodynamic equilibrium in each of subsystems is $\sim 10^{-14}$ s. The using of pico- and femtosecond laser pulses lets one to find new physical phenomena such as non-ideal and non-isothermal plasma generation of solid state density [5, 6]. The results presented in [5, 6] was explained by generation of strongly non-equilibrium states characterized by the production of non-equilibrium phonons in a metal (ion-acoustic turbulence) (see [7, 8]). As electrons have a mass and an electric charge their ultrashort beams can excite highly non-equilibrium states in matter at relatively lower intensities in comparison with laser ones (see [7], in which was established that significant feature of the action of high-power electron pulses of $10^{-14} < \tau_b < 10^{-9}$ s duration is the high rate of the matter deformation).

The aim of the paper proposed is to study experimentally and theoretically the characteristics of the picosecond electron beam produced at different interelectrode gap lengths and for various cathode materials, as well as the study of the rear side dynamics of the metal target under the action of such beam by means of an interferometric technique. Taking into account the results of mathematical modeling of wave processes in the continuum approximation, this results allows one, in our opinion, to study the slow relaxation of a non-equilibrium states excited by such beam.

## 2. Experimental setup

The experiments were carried out on EXCITOR setup shown on Fig. 1a. It was designed on the basis of "RADAN 300" compact high voltage generator with double-pulse-forming line and resonant Tesla transformer [10] (this generator is marked by "1" on figure 1). The coaxial oil-filled line with length of 30 cm and impedance of $Z_w$= 50 Ω with adjustable peaking and chopping discharge gaps filled with nitrogen at pressure of 40 atm were used to cut and to reduce generator voltage pulses. Electron beam was formed in a vacuum chamber at pressure of $10^{-6}$ Torr (2, figure 1). The left insert in the top of figure 1a shows the interelectrode gap.

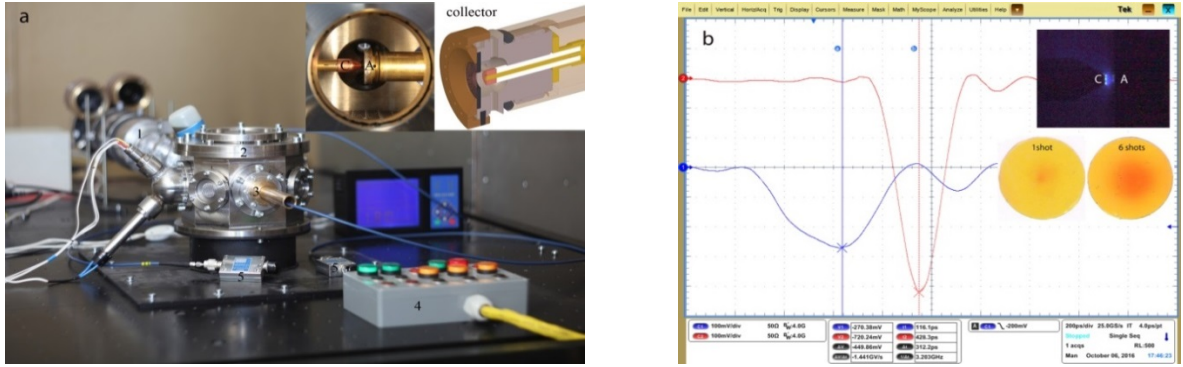

Figure 1. a) View of the experimental EXCITOR setup: 1 – generator RADAN-300 [10] output; 2 – vacuum chamber; 3 – output of current collector (on the insert: view of interelectrode gap (left) and drawing of the current collector cross-section (right)); 4 – vacuum chamber remote control; 5 –high voltage attenuator 142-NMFP-20B Barth Electronics, Inc. (USA); b) Typical voltage waveform recorder on the input of oil-filled impedance matching line with $Z_w = 50\Omega$ (C1) and current waveform on the collector output with $Z_w = 50\Omega$ (C2). On inserts: upper image is view of discharge glow in optical spectral range (cooper cathode, $L_{gap}$ = 2 mm); two lower images are the electron beam prints on radiochromic film.

The voltage at the beginning of the matching line was measured using a capacitive divider. The beam current after passing through two copper foils with a total thickness of $36\,\mu m$ was measured by a current collector based on a Faraday cup with wave impedance of $Z_w = 50\Omega$ (on figure 1a the collector output marked by "3"; the drawing of the collector cross section is shown on the right insert). Signals from the capacitor divider and the current collector were transmitted by cables with the impedance of $Z_w = 50\Omega$ and passband of $H_f = 18$ GHz to the inputs of a four-channel digital oscilloscope DPO70404C with the passband of $H_f = 4$ GHz. Barth 142-NMFP-20B attenuators with $\Delta_s = 20$ dB class 2.5 kV / 400 ns, the impedance of $Z_w = 50\Omega$ and passband of $H_f = 32$ GHz (marked by "5" on figure 1a) and AEROFLEX 23-6-34 ($\Delta_s = 6$ dB) and 23-20-34 attenuators with $\Delta_s = 20$ dB class 1 kW / 1 μs, $Z_w = 50\Omega$ and $H_f = 18$ GHz were used to reduce the signal amplitude. The figure 1b shows typical waveforms of the voltage at the input of the matching line and the current at the output of the current collector. The view of discharge glow of copper cathode with interelectrode gap $L_{gap}$ = 2 mm and electron beam prints on radiochromic film are shown on insert of figure 1b. The intensity of the film blackening can be approximated by normal distribution:

$$I(x) = \frac{A}{\sqrt{2\pi}} \exp\left(-\frac{(x-x_0)^2}{2\sigma^2}\right)$$

with parameters: $A = 11.17$, $x_0 = 4.2$ mm, $\sigma = 0.81$ mm for a single shot and $A = 22.74$, $x_0 = 4.25$ mm, $\sigma = 1.88$ mm for six shots, respectively. Diameters of exposition spot at half maximum were $FWHM_1 = 1.9$ mm and $FWHM_6 = 4.42$ mm, respectively.

Cathodes of tungsten, graphite, copper and samarium were used for measurements of picosecond electron beams characteristics. They had the same shape and dimensions (see left insert on figure 1a).The cathodes have a small recess at their ends so the sharp edge was formed at distance of $R_c = 1$ mm from cathode axis. One can see glowing cathode spots on it (the upper insert of figurer 1b.)

The figure 2a schematically shows the part of experimental setup by which measurement of displacement of the rear side of the anode was performing. Michelson interferometer scheme in which the anode rear side was one of the mirrors was used in our experiments.

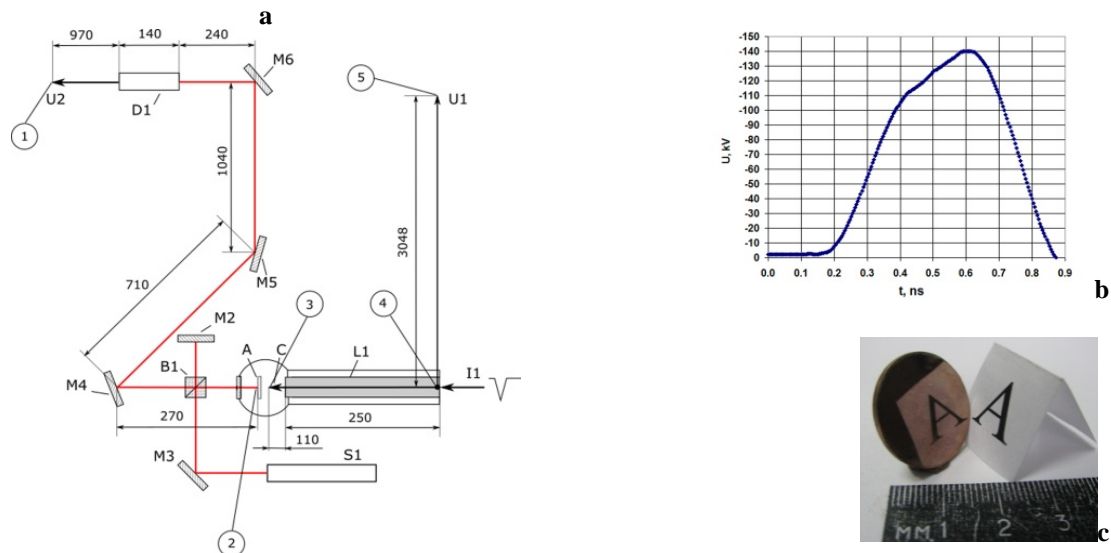

Figure 2: a) The scheme of the interferometer part. The figure shows the dimensions in mm for delays calculating. 1 is the "U2" input of the oscilloscope, 2 is the anode, 3 is the cathode, 4 is the voltage divider, 5 is the "U1" input of the oscilloscope; b) electron beam voltage waveform; c) rear side of the anode which is mirror of the interferometer.

The typical waveform of the high-voltage pulse *I1* is shown on figure 2b. It was transmitted to the input of the *L1* matching line, (figure 2 a) and also to the voltage divider (is not shown on the figure) and then to the *U1* input of the oscilloscope (see figure 2a). In this case the signal delay between the voltage divider and the oscilloscope input was about 13.1 ns. The pulse was transmitted to the cathode ("C" in figure 2a) by the matching line. The total signal delay in the *L1* matching line was 1.3 ns. The voltage pulse on the cathode forms an electron beam in the vacuum chamber. This beam enters into the anode («*A*» in figure 2a). The last one was made of copper plate with a thickness of 0.1 or 3 mm. Anode rear side was polished (figure 2c) to be used as one of the mirrors of the Michelson interferometer. The last one also includes *M2* mirror and *B1* beam splitter cube. The radiation of *S1* laser forms an interference image shown on figure 3.

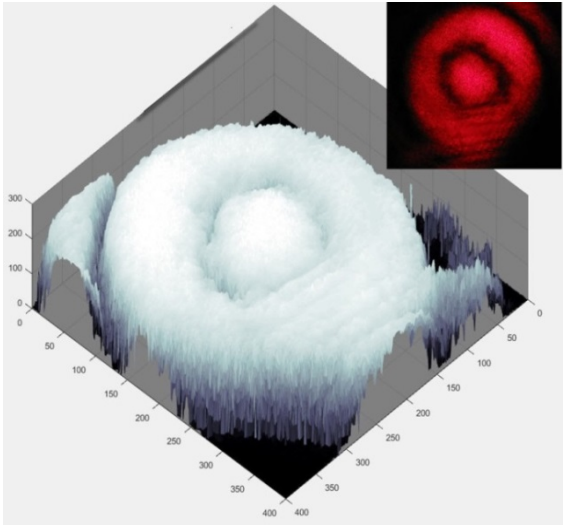

Figure. 3. Interference image and the intensity profile.

## 3. Experimental characteristics of picosecond electron beams

The length of the interelectrode gap was varied from 0.5 to 6.1 mm in order to find the current and energy of the electron beam corresponding to the maximum current pulse amplitude. Figure 4 shows the results of these experiments. It represents the amplitude (figure 4a) and the half-height duration (figure 4b) of the voltage pulse, and also the amplitude (figure 4c) and the half-height duration (figure 4d) of the electron beam current pulse in dependence on the length of the diode gap.

The full electromagnetic relativistic particle-in-cell (PiC) KARAT code was used to reconstruct the electron energy, as well as the shape and amplitude of the beam current pulse at the front side of the anode foil. In this case, the large-particle technique to simulate the dynamics of electrons beam in electromagnetic field was used [11]. The geometry of the calculated domain fully corresponded to the geometry of the real vacuum diode and the current collector. The large particles technique was used to simulate the electrons emitted from the cathode tip area according to the Richardson law [12] at a temperature of 293 K. The initial particle distribution $f(\varepsilon,\varphi,\theta)$ was randomly assigned in the energy and angular coordinates $\varepsilon \in (\varepsilon_{\min},\varepsilon_{\max}), \varphi \in (0,2\pi), \theta \in (0,\pi/2)$. In these calculations the minimum energy was assumed to be zero and the maximum one was chosen so that the collector current was close to the experimental current. This value of the initial energy varied from fractions - to ten electron-volts. The calculated voltage pulse at the input of the vacuum diode corresponded to the voltage first half-wave without account of the reflected waves since in the experiment the voltage pulse was measured at the beginning of the matching line having length of 30 cm (typical pulse shape is shown on figure 2b).

Figure 5 shows an examples of such calculations for a copper cathode with $L_{gap} = 2$ mm : figure 5a shows the voltage pulse at the diode input; figure 5b shows the current at the front side of the anode; figure 5c shows current on the electrode of the Faraday cup; figure 5d shows the current at the output of the current collector. One can see the wave processes in the coaxial line connecting the Faraday cup with the collector output cable smooth the fluctuations of the beam current. It is worth to note the calculated energy spectra of the beams are in good quantitative agreement with the integral energy spectra measurements of subnanosecond electron beams, carried out in [13] with use of the Thomson spectrograph.

A similar method was used to simulate all the experiments, the results of which are shown in figure 4. With the use of such analysis and comparison of the experiments and simulation results, optimal for the maximum values of the pulse amplitude of the current and energy of the beam electrons, as well as the minimum values of its duration were evaluated (see Table 1). Optimal values of the beam parameters correspond to the lengths of the interelectrode gap in the range of (1.0-3.1) mm (see Table 1).

Among the cathode materials used the best results were obtained with using of graphite cathode (see Table 1). In this case, the maximum amplitude of the beam current pulse, the smallest values of its width and the maximum value of the electron energy corresponding to their energy distribution were obtained.

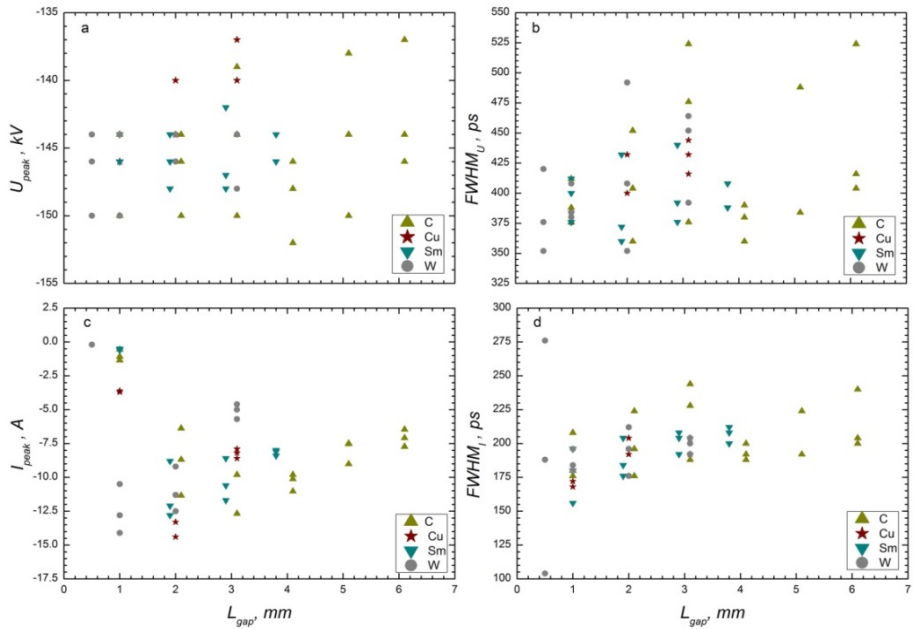

Figure 4. Dependencies on the length of the interelectrode gap $L_{gap}$: **a, b)** the amplitude $U_{peak}$ and duration of the voltage pulse at half-height $FWHM_U$, respectively, **c, d)** the amplitude $I_{peak}$ and duration of the current pulse at half-height $FWHM_I$

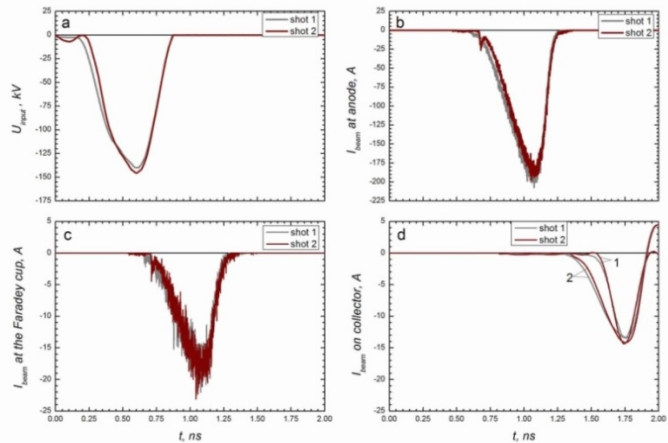

Figure 5. Example of the inverse task solving for currents restoration on the front anode side and on the electrode of Faraday cup: a) voltage pulse; b) current on the anode; c) current on the electrode of Faraday cup; d) current on the collector output. Curves 1 show the experimental results and curves 2 –the data of calculation.

**Table 1 Optimal characteristics of picosecond beams**

| Cathode material | Interelectrode gap $L_{gap}$, mm | Experiment | | | | Calculation | | |
|---|---|---|---|---|---|---|---|---|
| | | Voltage pulse | | Current pulse on the output of collector | | Current pulse on the front side of anode | | Amplitude of mean electron energy |
| | | $U_{peak}$ kV | FWHM ps | $I_{peak}$ A | FWHM ps | $I_{peak}$ A | FWHM ps | $\langle W \rangle$ keV |
| Cu | 2.0 | $-(142\pm2)$ | $416\pm16$ | $-(13.85\pm0.55)$ | $198\pm6$ | -175 | 321 | $213.73\pm6.54$ |
| Sm | 1.9 | $-(146\pm2)$ | $388^{+44}_{-28}$ | $-(11.233^{+1.567}_{-2.433})$ | $188^{+16}_{-12}$ | -132 | 330 | $234.96^{+3.13}_{-5.42}$ |
| W | 1.0 | $-(148^{+2}_{-4})$ | $390^{+17.333}_{-10.667}$ | $-(12.467^{+1.913}_{-1.967})$ | $186.667^{+9.333}_{-6.667}$ | -145 | 339 | $215.16^{+0.27}_{-0.15}$ |
| Graphite | 3.1 | $-(144.333^{+5.667}_{-5.333})$ | $458.667^{+65.333}_{-82.667}$ | $-(10.757^{+1.913}_{-0.957})$ | $220^{+24}_{-32}$ | -234 | 248 | $238.63^{+14.99}_{-18.53}$ |

It worth to note that the copper cathode is not worse than the graphite one because the value of $\langle W \rangle = 213\pm6.54$ keV was obtained for less values of the interelectrode gap $L_{gap} = 2$ mm and the voltage pulse amplitude $U_{peak} = -(142\pm2)$ kV. That is why the copper cathode and anode were used during displacement investigations of the anode rear side. This investigations were carried out for the interelectrode gap equals to 2 mm.

## 4. Dynamics of the target rear surface displacement: experiment and simulation

Only central fringe of the interference pattern (figure 3) was used for anode rear side displacement evaluation. This central fringe was cut out by diaphragm. One can see on figure 6 this fringe marked by the white circle. The light flow was recorded by photodiode (see figure 2a, D1) and digital oscilloscope Tektronix DPO70404C with passband of $H_f = 4$ GHz. The typical oscillograms for the different anode thicknesses are shown in figure 7.

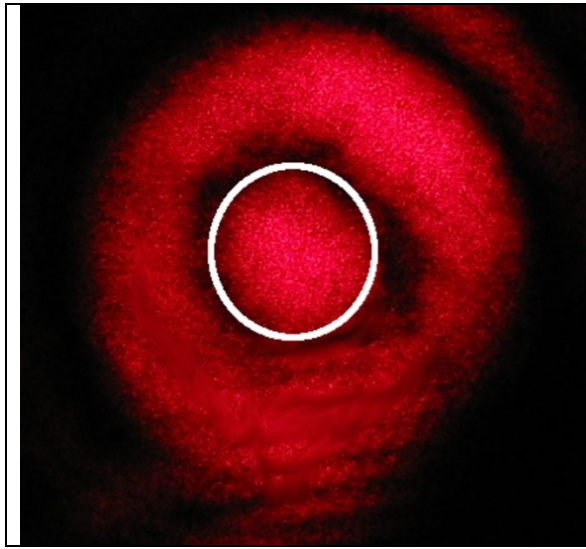

Figure 6. The central fringe of the interference image (marked by the white circle).

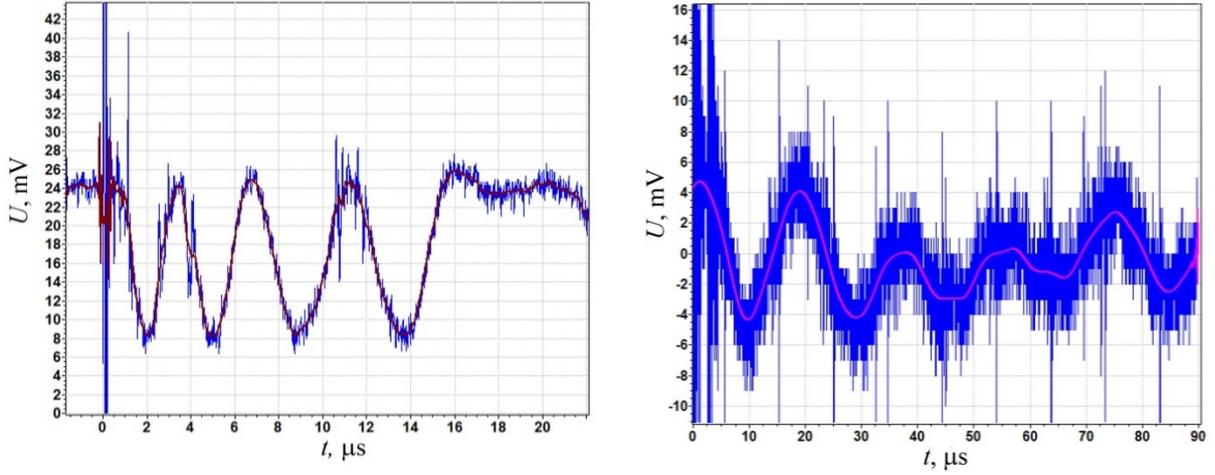

Figure 7. The typical photodiode (D1, Fig. 2a) signals at different anode thickness: the left image shows results for the 0.1 mm and the right - 3mm).

It is common knowledge the interference signal is described by the following equation

$$I = I_0 + I_s \cos\left(\frac{2\pi\Delta}{\lambda}\right),$$

where $I_0, I_s$ are the amplitudes of constant and variable components, respectively, $\Delta$ is the optical path difference, $\lambda$ is the laser wave length (in our case $\lambda$ = 632.8 nm). The difference in arms length of the interferometer is $x = \Delta/2$. For example, if the "A" mirror (see figure 2a) shift is $\lambda/2$ (i.e. length difference of interferometer arms change), then the optical path difference will be changed on $\lambda$ and the phase difference change will be $2\pi$.

The displacement of the rear side of the "A" mirror occurs as a result of the action of a picosecond electron beam with the parameters presented in Section 3. This displacement is a result of wave processes occurring inside the anode due to the relaxation of the non-equilibrium states excited by such electron beams. Therefore, it can be described by the following relation:

$$x(t) = x_0 + x_1 \sin\left(\frac{\pi(t-t_0)}{T}\right), \qquad (1)$$

where $x_0$ is the initial optical path difference, $t_0$ is the delay time of the displacement beginning of the anode rear surface relatively electron beam action, $x_1$ is displacement amplitude, $T$ is displacement period. Then the change in the signal of the photodiode will be determined by the following relation:

$$I = I_0 + I_s \cos\left(\frac{4\pi}{\lambda}\left(x_0 + x_1 \sin\left(\frac{\pi(t-t_0)}{T}\right)\right)\right). \qquad (2)$$

Figure 8 shows the approximation of the photodiode signal (shown on left side of figure 7) by using relation (2) for the anode thickness of 0.1 mm. The following approximation results were obtained: $x_0$ = -=1.30·10$^5$ nm; $x_1$ = 1.32·10$^5$ nm; $t_0$= -259.06, ns; $T$=575.11ns. The photodiode signal obtained for the anode with a thickness of 3 mm is approximated in a similar.

Figure 9 shows the results of processing the interferometric results, and figure 10 shows the speed estimation of the anode rear side. As one can see on the insert of figure 9 the rear anode side with thickness of 3 mm begins to move with a time delay of $t_d$=1.4 µs. However, for an anode thickness of 0.1 mm this time delay is poorly resolved in the presented time scale. To resolve it the results obtained with a sampling of 0.4 ns per sample was used (figure 11). Figure 11 shows the displacement of the anode rear side with thickness of 0.1 mm is stably recorded with delay of $t_d = 225 \pm 30$ ns.

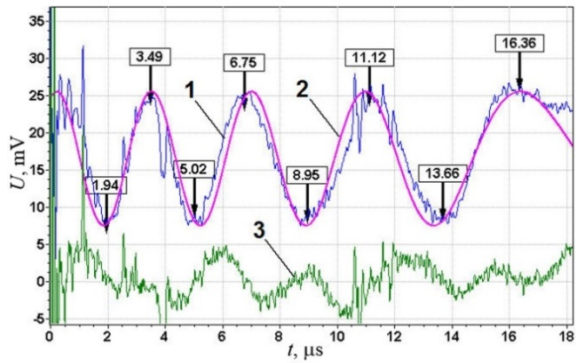

Figure 8. Example of the approximation of the photodiode signal shown on the left side of figure 7: 1) experimental data, 2) approximation result, 3) difference between the experimental and approximation data.

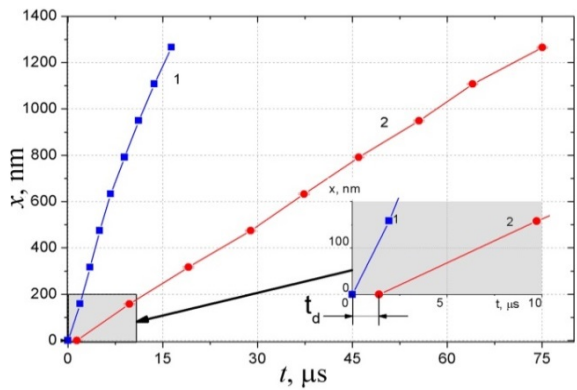

Figure 9. The time dependencies of the displacement of the central part of the rear side of the anode with the thickness of 1) 0.1 mm, 2) 3 mm. Here $t_d$ is the displacement delay time

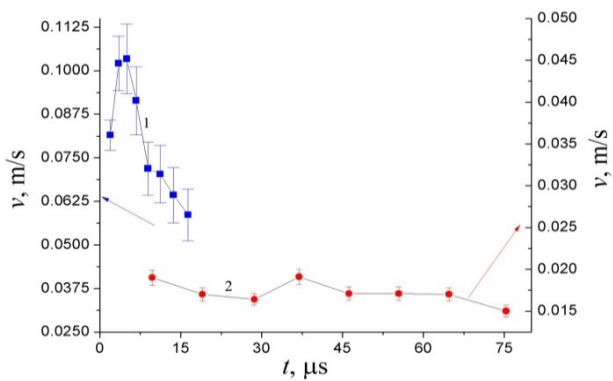

Figure 10. The speed estimation for the central part of the rear side of the anode. The notations for the curves are the same as on figure 9

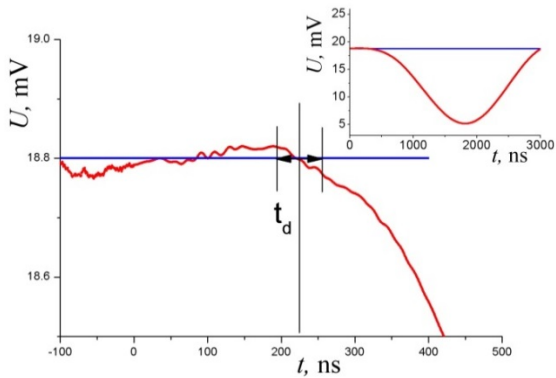

Figure 11. The waveform of the photodiode signal for the anode thickness of 0.1 mm.

This delay is in a good agreement with the parameter of -$t_0$= 259.06, ns in relations (1) and (2) for an anode thickness of 0.1 mm.

The simulation was carried out within the framework of software realizing the solution of the equations of two-dimensional axisymmetric single-temperature mechanics of a continuous elastoplastic medium [14]. The purpose of this simulation was to evaluate the wave processes role occurring inside the anode caused by the action of the electron beam of $t_b = 0.5$ ns duration and the space-time evolution of the thermodynamic and mechanical anode matter characteristics.

The system of the equations includes equations for the energy release calculation function in a matter irradiated by the beam; movement and energy balance equations; wide-range state equation and also equations for the stress tensor of a continuous medium. Calculation of the energy release function in matter under the fast electrons beam action is based on the solution of the corresponding kinetic equations [15]. Usually, the elastoplastic flow model which is based on the von Mises plasticity model [17] to describe the characteristics evolution of solids under pulsed action is used [16]. In our software we have used more detailed in comparison with von Mises model [17] description of plastic flows within the Prandtl-Reuss model [18, 19, 20] in which the plastic deformations velocities tensor is used directly. To solve the equations system for the continuum mechanics the numerical-analytic method proposed in [21] was used. Since the duration of the electron beam irradiation in our experiments was only 0.5 ns, we used the simplest Mie–Gruneisen state equation.

The calculated area had a radius of 3.5 mm. The target thicknesses were equal to 0.1 and 3 mm in the calculations. The beam parameters and the current density distribution along the radius were set according to the optimal beam characteristics (in our case for the copper anode) determined in Section 3. The total beam energy was $1.265 \cdot 10^{-2}$ J. The current density distribution was determined in the Gaussian form with a dispersion parameter of $\sigma = 0.81$ mm according to prints on the radiochromic film of the electron beam (see Section 2).

The numerical results show the temperature of the anode material increases only for few degrees in comparison with the initial temperature ($T_0 = 300$ K). At the same time, the mechanical stresses reached tens MPa, but not reaching of 68.5 MPa yield strength point for copper. The calculated displacement of the anode rear side with a thickness of 3 mm did not exceed 10 nm, which is 2 orders of magnitude less than the one observed in the experiment. In our opinion, this is due to ignoring of the interaction peculiarities of the electron beam with duration less than 1 ns with the target. The main of these peculiarities is defined by fact that during the irradiation the beam charge introduced into the anode with almost motionless ions induces in it electric (electronic) currents. The interaction of these currents with the beam main current leads to a much more intense mechanical action on the anode matter[1].

## 5. Conclusion

Thus, in this paper: (1) the optimal (from the point of view of the maximum values of the current amplitude and the average energy of the beam particles with the minimum duration of the beam pulse) characteristics of the picosecond electron beams were determined both experimentally and by numerical simulation within the framework of the fully electromagnetic relativistic Particle-in-Cell (PiC) KARAT code [11] to excite highly non-equilibrium states in the irradiated metal target; (2) an interferometric technique of recording of the anode rear side displacement as a result of the irradiation by the picosecond electron beam was implemented. In case of the appropriate modification of the theoretical model this technique will allow one within the framework of the continuum mechanics to investigate the slow spatiotemporal evolution of the non-equilibrium states of a metal excited by such irradiation.

---

[1]In separate work a theoretical study of this, in the general case, nonlinear phenomenon within the framework of the metal model proposed in [22] is planned. It is also planned to carry out with use of this interferometric technique experimental studies dynamics absorption of pulsed laser radiation with accent on the first nanosecond of processes the cathode-facing side of the anode irradiated by the picosecond electron beam.

**Acknowledgments**

This work was performed as part of State Task No. 0389-2015-0023, and under the partial financial support of the RFBR in terms of an optical measuring of the surface velocity of the target and the results obtained with its help (project No. 16-08-00466), as well as the Ural Branch of RAS within the UB RAS complex program for fundamental investigations (project No. 18-2-2-15).